\documentclass{ws-procs9x6}
\usepackage{graphicx}
\usepackage{bm}

\newif\ifpreprint
\preprinttrue                  

\def\alphas{\alpha_{\rm s}}
\def\Qs{Q_{\rm s}}
\def\x{{\bm x}}
\def\p{{\bm p}}
\def\k{{\bm k}}
\def\A{{\bm A}}
\def\Mrowczynski{Mr\'owczy\'nski}

\begin{document}

\title{
\ifpreprint
  Quark-Gluon Plasma
\fi
  Thermalization and Plasma Instabilities%
}

\author{Peter Arnold}

\address{
    Department of Physics,
    University of Virginia \\
    P.O. Box 400714,
    Charlottesville, VA 22901-4714 \\
}

\maketitle

\abstracts{
In this talk, I review the important role played by plasma instabilities
in the thermalization of quark-gluon plasmas at very high energy.
\ifpreprint%
  [Conference talk presented at
  Strong and Electroweak Mattter 2004,
  Helsinki, Finland, June 16--19.]
\fi
}

\section{Introduction}

Here is a basic question: What is the (local) thermalization time for
quark-gluon plasmas (QGPs) in heavy ion collisions?  That's a difficult
question, so let's ask a simpler one: What is it for {\em arbitrarily high
energy}\/ collisions, where the running coupling constant is small,
$\alphas \ll 1$?  Even that turns out to be complicated, so let me
focus on an even simpler version, first posed by Baier, Mueller, Schiff
and Son:\cite{BMSS} How does that time depend on $\alphas$?
In particular, in the saturation picture of heavy-ion collisions,
what is the exponent $n$ in the relation
\begin {equation}
   t_{\rm eq} \sim \frac{\alphas^n}{\Qs} \,?
\end {equation}

\begin{figure}[ht]
\label {fig:TRACE}
\begin{center}
\ifpreprint
  \includegraphics[scale=0.40,angle=-90]{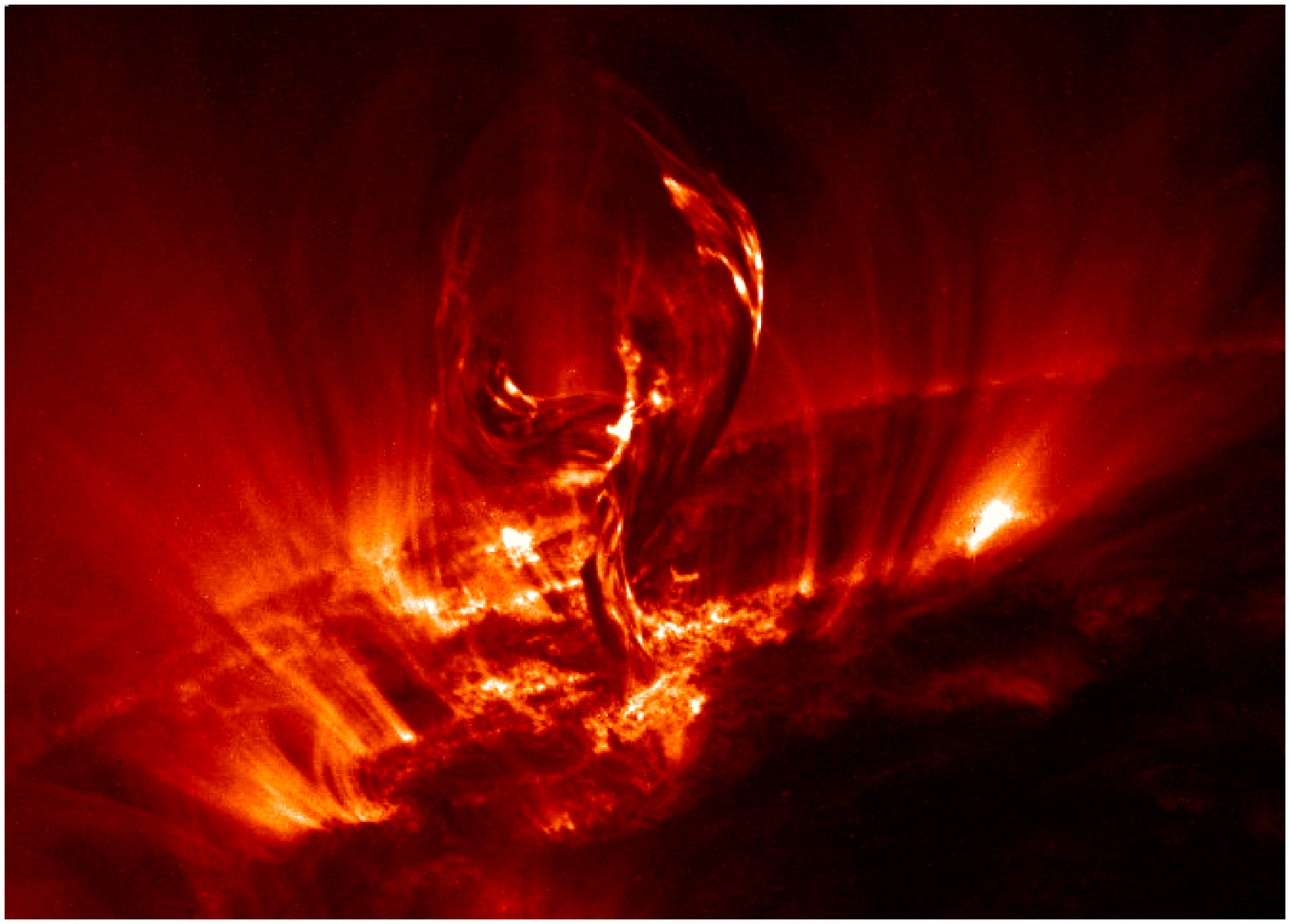}
\else
  \includegraphics[scale=0.40,angle=-90]{filament_trace_big_BW.eps}
\fi
\end{center}
\caption{%
  Image of a solar coronal filament from NASA's TRACE satellite,
  from
  $\langle${\sf http://antwrp.gsfc.nasa.gov/apod/ap000809.html}$\rangle$.
\ifpreprint
\else
  (It's much more dramatic in color---see the web page.)
\fi
}
\end{figure}

Before discussing this question in more detail, I wish to make a general
side comment about plasma physics.  Plasma physics is complicated!
This is made abundantly clear simply by looking at pictures of various
plasma phenomena, such as the image in Fig.\ \ref{fig:TRACE}
of a solar coronal filament from NASA's TRACE satellite.
Theoretical discussions of quark-gluon plasmas, however, are generally
much less complicated.  There are several reasons why such discussions
can usually avoid the full complication of traditional plasma physics.
First, much of the theoretical literature discusses QGPs that are
at or very close to thermal equilibrium.  The physics of plasmas near global
thermal equilibrium is much less complicated than the physics of
non-equilibrium situations.  Because electromagnetic interactions
are long-ranged, traditional electromagnetic plasmas can be very complicated
even when they are in {\it local}\/ thermal equilibrium.  Macroscopic
currents in one region of the plasma can interact magnetically with
other currents in other regions, over tremendous distance scales, creating
complicated structures like Fig.\ \ref{fig:TRACE}.  Non-Abelian plasmas,
however, are somewhat different.  From theoretical studies of the
equilibrium properties of such plasmas, we know that the non-Abelian
interactions cause magnetic {\it confinement}\/ over distances of order
$1/(g^2 T)$.  It is reasonable to assume that, even dynamically,
color magnetic fields cannot exists on distance scales larger than the
confinement length.  So, unlike traditional electromagnetic plasmas,
there are no large-distance magnetic fields.  As far as the color
degrees of freedom are concerned, the long-distance effective theory of a
non-Abelian plasma is hydrodynamics rather than magneto-hydrodynamics.

The color magnetic fields {\it can}\/
play a role on small scales.
But the full complications of plasma physics might be ignored
on small distance scales if
the relevant physics on those scales is weakly interacting.
This was the proposal of the original bottom-up scenario for thermalization
of quark-gluon plasmas\rlap.\cite{BMSS}  However, as we shall discuss, even
at small distance scales, there can be plasma instabilities.
These instabilities cause the growth of non-perturbatively large magnetic
fields, bringing
in some of the complicated non-linear physics of traditional plasma
physics.

\begin{figure}[ht]
\label {fig:expansion}
\begin{center}
\ifpreprint
  \includegraphics[scale=0.15]{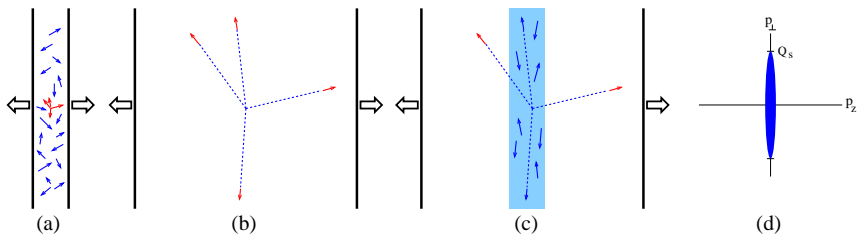}
\else
  \includegraphics[scale=0.15]{expansion_BW.eps}
\fi
\end{center}
\caption{%
   Approximately free expansion at relatively early stages of the bottom-up
   scenario.
}
\end{figure}

In the original bottom-up thermalization scenario of Baier {\it et al.},
the starting assumption is the saturation scenario, which assumes that
the formation of the quark-gluon plasma starts at times $t \sim \Qs^{-1}$
dominated by gluons
with momenta $p\sim\Qs$,
where the scale $\Qs$ is known as the saturation scale.
In this talk, these gluons will be referred to as ``hard'' gluons, since
we will soon be discussing even softer momentum scales.
This situation is depicted in Fig.\ \ref{fig:expansion}a.
The occupation numbers of each gluon mode are initially non-perturbative,
with $f(\x,\p) \sim 1/\alphas$, where $f$ is the phase-space density.
As the system expands 1-dimensionally immediately after the collision,
the density per unit volume decreases, and one might therefore expect
the hard gluon interactions to become more perturbative.  To understand
what happens next, let's ignore these perturbative interactions for
the moment and think about free expansion.  As the nuclear pancakes
fly apart after the collision, the gluons, which started in the center,
will separate themselves in space according to the $z$ components of their
velocities, as $\Delta z \sim v_z t$.  This is depicted in
Fig.\ \ref{fig:expansion}b.  The gluons left in the central region will
be those with small $v_z$, as shown in Fig.\ \ref{fig:expansion}c.
As a result, the momentum distribution in that central region will have
an anisotropic pancake shape, as shown in Fig.\ \ref{fig:expansion}d.
In other regions, the momentum distribution is simply a boosted version
of this---that is, it looks the same if one works in the local rest frame.

It should be emphasized that this anisotropic momentum distribution has
nothing to do with the usual elliptic flow distribution measured in
non-central heavy ion collisions.  Fig.\ \ref{fig:expansion}d is a statement
of the local momentum distribution in the local fluid rest frame
at very early times, before collisions
have brought the system to local equilibrium.  Elliptic flow is instead
a measure of the {\it net}\/ fluid flow on large scales of a system that has
quite possibly come to local thermal equilibrium (and so locally has
isotropic momentum distributions in the local fluid rest frame).

\begin{figure}[ht]
\label {fig:cascade}
\begin{center}
\ifpreprint
  \includegraphics[scale=0.40]{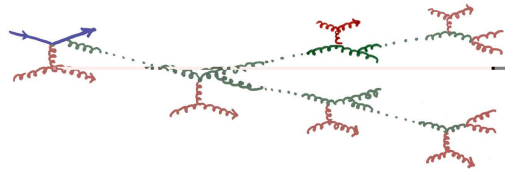}
\else
  \includegraphics[scale=0.40,angle=-90]{cascade_BW.eps}
\fi
\end{center}
\caption{%
   Cascading process for energy transfer from hard particles to soft sector
   in the original bottom-up scenario.
   Straight lines denote hard gluons, $p \sim \Qs$.
}
\end{figure}

In the original bottom-up scenario, equilibration of the plasma was
assumed to occur through individual 2-body collisions between particles
(with some LPM effect thrown in).  In the first stage of the scenario,
$1 \ll \Qs t \ll \alpha^{-5/2}$, the important processes were small
angle scattering, which slightly widens the hard particle distribution
in $p_z$, and soft Bremsstrahlung from colliding hard particles,
which creates soft gluons with momenta $k \ll \Qs$.  In the second
stage, $\alpha^{-5/2} \ll \Qs t \ll \alpha^{-13/5}$, these soft gluons
come to dominate the number density of particles, but the hard gluons
still dominate the energy density.  Collisions between the soft gluons
cause the soft gluons to thermalize.  Finally, the hard particles begin
to lose energy by Bremsstrahlung plus cascading, as shown in Fig.\
\ref{fig:cascade}.  In this scenario, local thermalization became complete
at $\Qs t \sim \alpha^{-13/5}$.

The original bottom-up scenario overlooked the possibility that
{\it collective}\/ processes (as opposed to sequences of individual collisions)
could play an important role in the equilibration of the plasma.
In the case at hand, these collective processes are related to the
appearance of plasma instabilities in the analysis of the equilibration
of the quark-gluon plasma.


\section{Plasma Instabilities}

The hero of this story is Stan
\Mrowczynski\rlap,\cite{mrow,mrow&thoma,randrup&mrow}
who over the years has been the
major proponent of the idea that plasma instabilities are important for
the equilibration of the quark-gluon plasma.  The application of this
idea to bottom-up thermalization was made by myself, Jonathan Lenaghan
and Guy Moore\rlap.\cite{alm}  A selection of other folks who have considered
the idea past and present include
Refs.\ \refcite{heinz_conf}--\refcite{birse}
and the work by Romatschke and Strickland\rlap,\cite{strickland} which
was reported on at this conference.

\begin{figure}[ht]
\label {fig:Pi}
\begin{center}
\ifpreprint
  \includegraphics[scale=0.40]{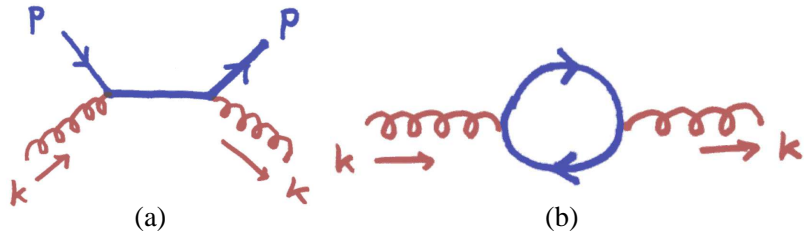}
\else
  \includegraphics[scale=0.40]{Pi_BW.eps}
\fi
\caption{%
  Self-energy of soft modes due to hard particles.
}
\end{center}
\end{figure}

Let me start with a slightly formal explanation of the origin of plasma
instabilities, and I will give a more physical picture afterward.
Imagine calculating the self-energy $\Pi(\omega,k)$ for a particle
moving through the plasma.  The self-energy represents the effect on
the particle of forward-scattering off of other particles in the medium,
as in Fig.\ \ref{fig:Pi}a, which one can alternatively calculate as
in Fig.\ \ref{fig:Pi}b using one's favorite formalism for
field theory in a medium, or more simply calculate using linearized
kinetic theory.
(Here, the straight lines represent hard particles,
which are gluons in the bottom-up scenario.)
The result is the same either way.  Generically,
if the momentum distribution $f(\p)$ of hard particles is anisotropic,%
\footnote{
   Here and throughout, I always assume $f$ is parity symmetric:
   $f(-\p) = f(\p)$.
}
one finds that there are negative eigenvalues of $\Pi^{ij}(0,\k)$.
(See, for example, the general arguments in Ref.\ \refcite{alm}.)  Such
negative eigenvalues indicate instabilities at small $k$, which means
exponentially growing soft gauge fields.

Let me give an analogy from scalar $\phi^4$ theory at finite temperature.
Imagine integrating out the hard particles to get an effective thermal
self-energy $\Pi$ of order $\lambda T^2$.  The effective linearized
equation of motion is then
\begin {equation}
   \omega^2 \phi = (k^2 + m_0^2 + \Pi)\phi \equiv (k^2 + m_{\rm eff}^2) \phi .
\end {equation}
Let's set $m_0 = 0$ to improve the analogy, since gluons do not have any
intrinsic mass.  If we were in a situation where $k^2+\Pi$ were less than
zero, then there would be solutions with $\omega=\pm i\Gamma$ pure
imaginary, which would lead to exponentially growing solutions to
the linearized equation.  Alternatively, think about the effective
potential,
\begin {equation}
   V_{\rm eff}(\phi) = m_{\rm eff}^2 \phi^2 + \lambda \phi^4
   = \Pi \phi^2 + \lambda \phi^4 .
\end {equation}
If $\Pi<0$, as happens in some multi-scalar theories\rlap,\cite{weinberg}
then the potential looks like a double-well potential: the naive
vacuum $\phi=0$ is unstable; there is exponential growth from $\phi=0$
of modes with $k < (-\Pi)^{1/2}$; and the growth stops once $\phi$ becomes
{\it non-perturbatively}\/ large.

\begin{figure}[ht]
\label {fig:weibel1}
\begin{center}
\ifpreprint
  \includegraphics[scale=0.40]{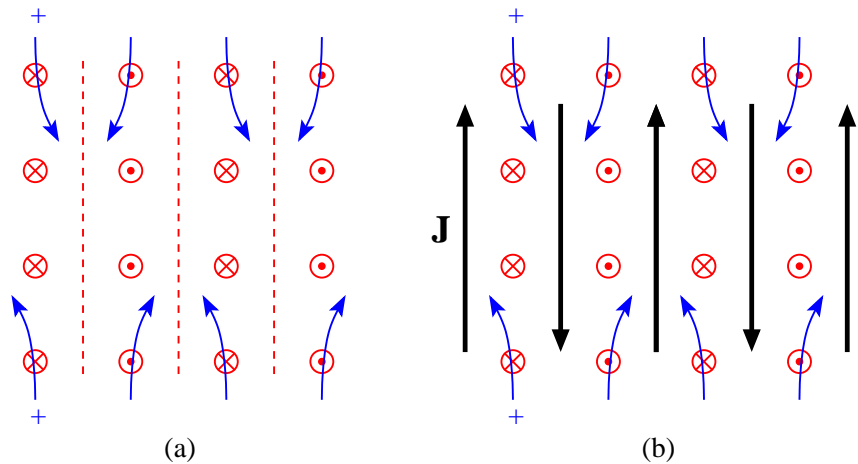}
\else
  \includegraphics[scale=0.40]{weibel1_BW.eps}
\fi
\caption{%
  Origin of Weibel instability.
}
\end{center}
\end{figure}

Now let's turn to a physical picture of the instability, which I will adapt
from Refs.\ \refcite{mrow},\refcite{alm}.
For simplicity, imagine two inter-penetrating,
homogeneous streams of charged hard particles, one going up the page and one
going down, which I'll call the $\pm x$ directions.  Now also imagine that,
due to some fluctuation, there is a very tiny seed magnetic field of
the form ${\bm B} = B {\bm e}_y \cos(kz)$, as shown in
Fig.\ \ref{fig:weibel1}a.  Here, crosses denote magnetic fields pointing
into the page, and dots fields pointing out of the page.  Using the
right-hand rule, you can check that the magnetic fields bend the trajectories
of positively charged particles in the directions shown.  This then focuses
the net downward and upward currents into different channels, as shown in
Fig.\ \ref{fig:weibel1}b.  Again using the right-hand rule, one finds that
these currents in turn create magnetic fields that {\it add}\/ to the
original seed field.  With bigger fields, the effect becomes more
pronounced, and the fields continue to grow through this mechanism.
This is the Weibel instability.

\begin{figure}[ht]
\label {fig:weibel2}
\begin{center}
\ifpreprint
  \includegraphics[scale=0.35]{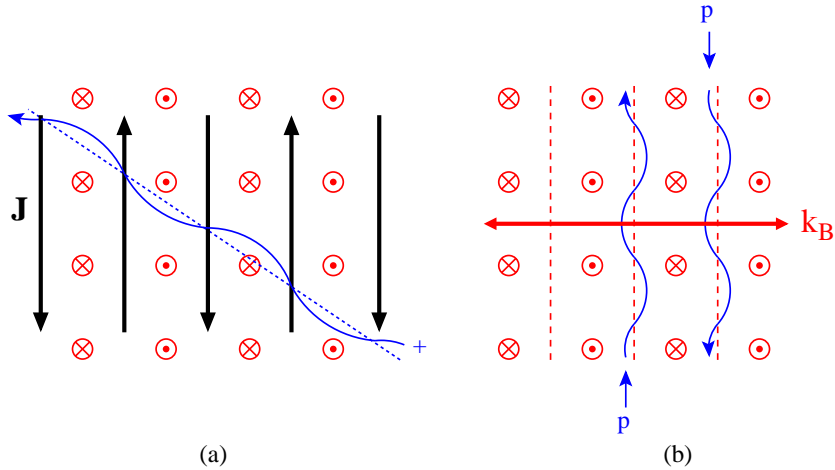}
\else
  \includegraphics[scale=0.35]{weibel2_BW.eps}
\fi
\caption{%
  (a) untrapped particles; (b) trapped particles.
}
\end{center}
\end{figure}

We get a seemingly contradictory picture of what happens if we
considers hard particles which move in other directions, such as
in Fig.\ \ref{fig:weibel2}.  Following how these particles are
by the seed magnetic field, we find they are directed a little more
upward in some regions and a little more downward in others.
This results in a net current as shown in the picture, but these
currents create magnetic fields which {\it oppose}\/ the seed field.
Let $\p$ be the original momentum of the hard particle and $\k$ the
wave number of the soft magnetic field fluctuation.  What happens is
that particles with $\p\cdot\k \simeq 0$ get trapped in
valleys, as shown in Fig.\ \ref{fig:weibel2}b, and give the
de-stabilizing effect discussed earlier.  Other particles, with
$\p\cdot\k \not\simeq 0$, are ``untrapped'' as in
 Fig.\ \ref{fig:weibel2}a, and give a stabilizing effect.
For {\it isotropic}\/ $f(\p)$, these two contributions turn out to
cancel, giving $\Pi^{ij}(0,\k)=0$.

Now, instead of an isotropic hard particle distribution $f(\p)$, think
of the one depicted earlier in Fig.\ \ref{fig:expansion}d.
For $\k$ in the $z$ direction, we will get a relatively smaller
percentage of particles with $\p \cdot \k \simeq 0$ than we would
with an isotropic distribution, and so the net effect will be stabilizing.
On the other hand, for $\k$ in the $\perp$ direction, we will get a
relatively larger percentage of particles with $\p\cdot\k \simeq 0$
than we would with an isotropic distribution, and so the net effect
will be de-stabilizing.  For $\k$ at significant angles to the axis,
most of the particles with have $\p\cdot\k \not\simeq 0$, and we will
tend to get stability.  The moral of this story, to which we will later
return, is that the unstable
modes associated with this distribution
have $\k$ pointing very close to the $z$ axis.


\section{When does the growth stop?}

As in the scalar analogy discussed earlier, the growth of instabilities
should stop when the soft fields become non-perturbatively large.
From considering covariant derivatives $D = \partial - i g A$, the
effects of soft fields $A$ become non-perturbative when
$A \sim \partial/g$.  There are two possibilities for the momentum
scale associated with the derivatice $\partial$.  The first,
a possibility for both QCD and QED, is
that growth stops when the effects of the soft fields on {\it hard}\/
particles becomes non-perturbative, which will happen when
$A \sim p_{\rm hard}/g$ and corresponds to the trajectories of
hard particles being bent dramatically from straight lines.
The second possibility, which cannot occur in QED, is that growth stops
when the non-Abelian self-interactions of the soft fields become
non-perturbative.  This corresponds to $A \sim k_{\rm soft}/g$.
Whenever there is a significant separation $k \ll p$ between soft
and hard physics (as in the bottom-up scenario), these two possibilities
correspond to parametrically different scales for $A$.

Jonathan Lenaghan and I conjecture\cite{al} that growing QCD instabilities
``abelianize.''  That is, the growth stops when
$A \sim k_{\rm soft}/g \sim |\Pi|^{1/2}/g$, just as in QED.
That in turn suggests that the complicated stuff that happens afterward
is closely related to the mainstream plasma physics of (collisionless)
relativistic QED plasmas.  What follows is a summary of suggestive
arguments that we make for abelianization.

\begin{figure}[ht]
\label {fig:V}
\begin{center}
\ifpreprint
  \includegraphics[scale=0.35]{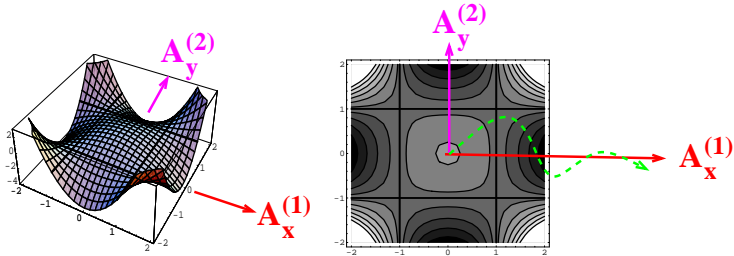}
\else
  \includegraphics[scale=0.35]{V_BW.eps}
\fi
\caption{%
  A picture of the effective potential.
}
\end{center}
\end{figure}

Start with the general HTL effective action for anisotropic $f(\p)$,
which I adapt from \Mrowczynski, Rebhan, and Strickland\rlap.\cite{mrs}
Shcematically,
\begin {equation}
   S_{\rm eff} = - \int_x F^2 - g^2 \int_{x\p} f(\p) \, W^2 ,
\label {eq:Seff}
\end {equation}
\begin {equation}
   W \equiv W_\alpha(x,\p) \equiv \frac{p^\mu}{p\cdot D} \, F_{\mu\alpha} .
\label {eq:W}
\end {equation}
Now imagine finding the effective potential $V_{\rm eff}$ by looking
at $S_{\rm eff}$ for time-independent configurations in $A_0{=}0$ gauge.
There is a problem, which is that $(p\cdot D)^{-1}$ is a complicated,
non-local operator.  But now recall that, for the hard particle
distribution of Fig.\ \ref{fig:expansion}d, the typical unstable modes have
$\k$ pointing very close the $z$ direction.  Inspired by this, let's
ignore $k_\perp$ altogether and consider configurations
$\A=\A(z)$ depending on $z$ only.
There is then an amazing simplification, noted by Blaizot and
Iancu:\cite{BIwaves}  $W$ given by (\ref{eq:W}) then turns out to be
{\it linear}\/ in $A$.  As a result, the HTL term $g^2 \int f W^2$
in the effective action (\ref{eq:Seff}) is then {\it quadratic} in $A$.
That means that it consists of nothing but the HTL self-energy, so that
\begin {equation}
   S_{\rm eff}[\A(z)] = - \int_x F^2 - \int_x A \Pi A .
\label {eq:Seff1}
\end {equation}
The $\Pi$ here is for $\k$'s proportional to $\hat z$.  It is also
transverse, corresponding to $\Pi_\perp(0,k\hat z)$.  This turns out
not to depend on the magnitude $k$ of $\k$.
I will define the constant $-\mu^2 \equiv \Pi_\perp(0,k\hat z)$.
The effective potential from (\ref{eq:Seff1}) is then
\begin {equation}
   V_{\rm eff}[A(z)] = B^2 - \mu^2 A_\perp^2 ,
\end {equation}
where $B$ is the full, non-Abelian magnetic field, which includes
non-linear terms in $A$.  Now consider the arbitrarily soft limit
$k \to 0$, where the derivative term in $B$ vanishes and only the
non-Abelian commutator term survives:
\begin {equation}
   V_{\rm eff} \to [A_x,A_y]^2 - \mu^2 (A_x^2+A_y^2) .
\end {equation}
What does this potential look like?  As an example, let's consider just
two of the degrees of freedom: color 1 of $A_x$ and color 2 of $A_y$.
A plot of the resulting potential is shown in Fig.\ \ref{fig:V}.
As one moves away from $A=0$, the potential bends down due to the
$-\mu^2 A^2$ term.  If both colors are present, it then later bends
back up again due to the quartic $[A_x,A_y]^2$ term.  However, if
the configuration is single-colored ($A_x^{(1)}=0$ or $A_y^{(2)}=0$),
the commutator vanishes, and the potential continues to run away.
If we were to roll a ball from the origin in such a potential, it would
eventually roll away down into one of the Abelian directions
$[A_x,A_y]=-0$.

The above argument suggests the following conjecture, which is
the conclusion of my talk: The growth of Weibel instabilities eventually
``abelianizes'' the soft gauge fields into the maximal Abelian subgroup
of the gauge group.  For SU(2) gauge theory, this would give traditional
U(1) plasma physics.  For SU(3) gauge theory, it would give U(1)$\times$U(1)
plasma physics, corresponding to two copies of Abelian electromagnetism.
Further discussion, including corroboration from numerical simulation results
of a related toy model, may be found in Ref.\ \refcite{al}.


\end{document}